\documentclass[10pt,journal,compsoc, onecolumn]{IEEEtran}
\usepackage{cite}
\usepackage[pdftex]{graphicx}
\DeclareGraphicsExtensions{.pdf,.jpeg,.png}
\usepackage{array}
\usepackage{amsmath}
\usepackage{amssymb}
\usepackage{mdwmath}
\usepackage{mdwtab}
\usepackage{eqparbox}
\usepackage{url}
\usepackage{hyperref}
\usepackage{multirow}
\usepackage{enumitem}
\usepackage{lineno} 
\usepackage[T1]{fontenc}

\begin{document}

\title{MedSAM2: Segment Anything in 3D Medical Images and Videos}

\author{Jun Ma$^*$, Zongxin Yang$^*$, Sumin Kim, Bihui Chen, Mohammed Baharoon, Adibvafa Fallahpour, \\Reza Asakereh, Hongwei Lyu, and Bo Wang$^\dagger$ 

\IEEEcompsocitemizethanks{
\IEEEcompsocthanksitem Jun Ma is with AI Collaborative Centre, University Health Network; Vector Institute, Toronto, Canada ($*$ Equal Contribution).
\IEEEcompsocthanksitem Zongxin Yang is with Department of Biomedical Informatics, Harvard Medical School, Harvard University, Boston, USA ($*$ Equal Contribution).
\IEEEcompsocthanksitem Sumin Kim is with Peter Munk Cardiac Centre, University Health Network; Department of Computer Science, University of Toronto; Vector Institute, Toronto, Canada.
\IEEEcompsocthanksitem Bihui Chen is with Peter Munk Cardiac Centre, University Health Network; Department of Computer Science, University of Toronto; Vector Institute, Toronto, Canada.
\IEEEcompsocthanksitem Mohammed Baharoon is with Department of Biomedical Informatics, Harvard Medical School, Harvard University, Boston, USA. Part of this work was done at the University of Toronto, Toronto, Canada.
\IEEEcompsocthanksitem Adibvafa Fallahpour is with Peter Munk Cardiac Centre, University Health Network; Department of Computer Science, University of Toronto; Vector Institute, Toronto, Canada.
\IEEEcompsocthanksitem Reza Asakereh participated in this project when he was with Peter Munk Cardiac Centre, University Health Network, Toronto, Canada.
\IEEEcompsocthanksitem Hongwei Lyu is with Peter Munk Cardiac Centre, University Health Network, Toronto, Canada. 
\IEEEcompsocthanksitem Bo Wang is with Peter Munk Cardiac Centre and AI Hub, University Health Network; Department of Laboratory Medicine and Pathobiology and Department of Computer Science, University of Toronto;  Vector Institute, Toronto, Canada($\dagger$Corresponding Author).
E-mail: bowang@vectorinstitute.ai}
}

\IEEEtitleabstractindextext{%
\begin{abstract}
Medical image and video segmentation is a critical task for precision medicine, which has witnessed considerable progress in developing task or modality-specific and generalist models for 2D images. However, there have been limited studies on building general-purpose models for 3D images and videos with comprehensive user studies. Here, we present MedSAM2, a promptable segmentation foundation model for 3D image and video segmentation. The model is developed by fine-tuning the Segment Anything Model 2 on a large medical dataset with over 455,000 3D image-mask pairs and 76,000 frames, outperforming previous models across a wide range of organs, lesions, and imaging modalities. Furthermore, we implement a human-in-the-loop pipeline to facilitate the creation of large-scale datasets resulting in, to the best of our knowledge, the most extensive user study to date, involving the annotation of 5,000 CT lesions, 3,984 liver MRI lesions, and 251,550 echocardiogram video frames, demonstrating that MedSAM2 can reduce manual costs by more than 85\%. MedSAM2 is also integrated into widely used platforms with user-friendly interfaces for local and cloud deployment, making it a practical tool for supporting efficient, scalable, and high-quality segmentation in both research and healthcare environments. 
\end{abstract}
}

\maketitle




\section*{Introduction}
Medical image segmentation plays a pivotal role in numerous clinical applications, including anatomical structure analysis~\cite{HeartNature20}, disease diagnosis~\cite{HeartMRI-NatMedicine}, surgery planning, and treatment monitoring~\cite{pancreasAI-NMed}. By delineating the boundaries of organs, lesions, and other relevant anatomies, segmentation algorithms provide clinicians with crucial information for precise disease analysis. Over the past decade, deep learning-based methods have revolutionized this field, delivering unprecedented performance on various segmentation tasks and benchmarks. For example, DeepLab~\cite{DeepLab}\cite{deeplabV3plus} has achieved human-level performance in left ventricle segmentation from echocardiography for ejection fraction assessment~\cite{HeartNature20}, which has proven to save time for both sonographers and cardiologists via blinding and randomization clinical trial~\cite{heartNature23}.
U-Net~\cite{UNet-MICCAI15} has been employed for accurate cell detection and segmentation in light microscopy images~\cite{UNet-NM18} and 3D nnU-Net~\cite{nnunet21} has been widely used in various anatomy and lesion segmentation, such as heart chamber segmentation in Magnetic Resonance Imaging (MRI) scans~\cite{HeartMRI-NatMedicine}, pancreas cancer and abdominal organ segmentation in Computed Tomograph (CT) scans~\cite{pancreasAI-NMed}\cite{FLARE22}, and whole-body lesion segmentation in Positron Emission Tomography (PET) scans~\cite{autoPET-NMI}. 

Driven by advanced network architectures~\cite{ViT2020ICLR} and large-scale datasets~\cite{SAM1-2023-ICCV}, recent trends in segmentation present a paradigm shift from specialist models tailored for specific tasks to generalist or foundation models capable of performing segmentation without extensive task-specific model development~\cite{FM-NaturePerspective,FM-ReviewYuting,FM-ReviewRuogu}.
One prominent example is the Segment Anything Model (SAM)~\cite{SAM1-2023-ICCV}, a pioneer segmentation foundation model in computer vision that has shown remarkable generalization ability across a wide range of two-dimensional (2D) natural image segmentation tasks. However, due to the substantial domain gap, its performance remains suboptimal in medical images~\cite{SAM1-Eval-Maciej}\cite{SAM1-Eval-XinYang}. Despite these limitations, SAM can be effectively adapted to the medical domain through transfer learning. For instance, models such as MedSAM~\cite{MedSAM} and SAM-Med~\cite{SAM-Med2D}\cite{SAM-Med3D} have demonstrated strong capabilities in segmenting various organs and abnormalities across diverse medical imaging modalities by fine-tuning SAM on large-scale medical datasets.

Despite the potential of these foundation models, their application to medical imaging is still limited and faces three main limitations. First, most medical image segmentation foundation models~\cite{MedSAM}\cite{SAM-Med2D} are primarily designed for 2D image data and may not capture the three-dimensional (3D) spatial relationships or temporal information in volumetric and video medical data. 
Second, although some studies have extended SAM to 3D image segmentation using 3D image encoders~\cite{SAM-Med3D} and adapters~\cite{Ma-sam,3DSAM-adapter,medsam-adapter} or developed interactive 3D segmentation models~\cite{segvol,VISTA3D,nnInteractive} to incorporate manual corrections, there is still a lack of general models to segment both 3D images and videos, which are frequently necessary in real-world clinical workflows. 
The state-of-the-art video segmentation model, SAM2~\cite{SAM2-2025-ICLR}, has shown great potential to fill this gap~\cite{sam2-medseg-survey,jundesam2,quanzhenglisam2,sam2-deployment}, but adaption on large-scale datasets has been underexplored.
Finally, large-scale validation of these models in practical image-labeling scenarios remains notably absent, leaving important questions about their scalability and utility in facilitating high-throughput medical image annotation tasks.

In this work, we address these limitations by presenting MedSAM2, a general model for 3D medical image and video segmentation. Specifically, we first curate a large-scale dataset consisting of more than 455,000 3D image–mask pairs and 76,000 annotated video frames, spanning multiple organs, pathologies, and imaging protocols for model development. Then, we build MedSAM2 by modifying and fine-tuning SAM2 on the large dataset. Extensive experiments show that MedSAM2 is capable of handling both volumetric medical scans and successive video frames, enabling versatile segmentation across diverse medical data. Furthermore, we conduct three user studies to demonstrate that MedSAM2 substantially facilitates annotation workflows for high-throughput and efficient segmentation, substantially reducing the time and effort required for creating large-scale medical datasets in various imaging modalities. MedSAM2 has the potential to transform clinical workflows by enabling more efficient diagnostic processes, treatment planning, and longitudinal monitoring across cardiology, oncology, and surgical specialties, where precise 3D organ and lesion segmentation is critical but traditionally time-consuming.

\begin{figure}[htbp]
\centering
\includegraphics[scale=0.23]{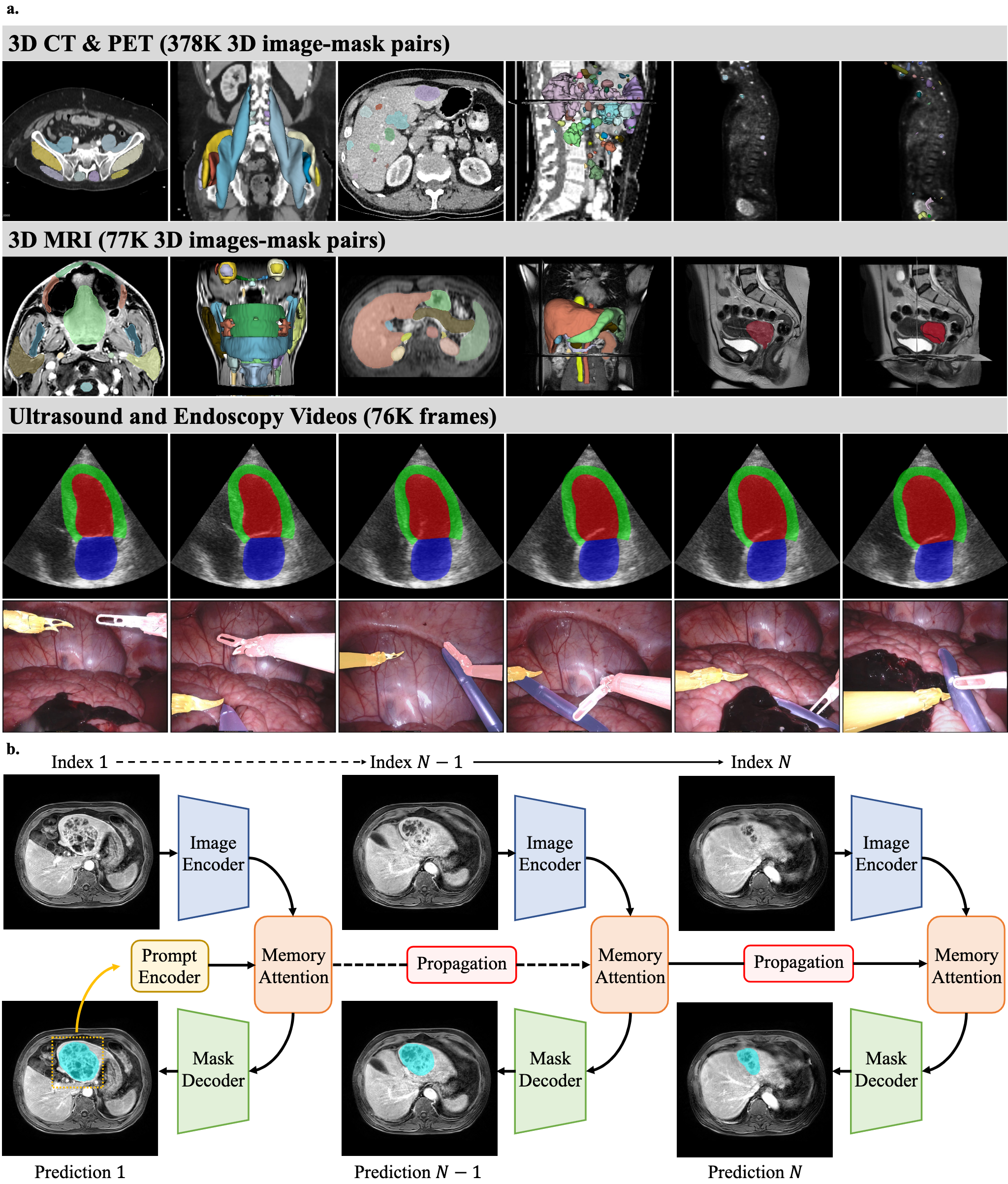}
\caption{\textbf{Dataset and network architecture for MedSAM2 development.} \textbf{a,} The dataset includes diverse 3D CT, PET, MRI images, ultrasound, and endoscopy videos. For each 3D image example, we visualize both 2D slices and 3D structures. For each video example, we visualize frames at different time points. 
\textbf{b,} MedSAM2 is a promptable segmentation network with an image encoder, a prompt encoder, a memory attention module, and a mask decoder. The image encoder extracts multiscale features from each frame or 2D slice. The memory attention module conditions the current frame features on past frames' features and predictions using streaming memory. The mask decoder generates accurate segmentation masks based on bounding box prompts and memory-conditioned features. This architecture enables MedSAM2 to effectively segment both 3D medical images and videos by exploiting spatial continuity across slices and frames.
}\label{fig:1}
\end{figure}

\section*{Results}
\subsection*{Dataset and model architecture} 
A large amount of training data is the foundation for developing generalist segmentation models. We assembled a large-scale and diverse 3D medical image and video dataset based on public datasets, including various normal anatomical structures and pathologies from various medical imaging modalities (Fig~\ref{fig:1}a, Methods, Supplementary Table 1).
In particular, we collected 363,161, 14,818, and 77,154 3D image-mask pairs for CT, PET, and MRI modalities, respectively. In addition, we curated 19,232 and 56,462 annotated frames for ultrasound and endoscopy, respectively. 

The pre-trained SAM2 model~\cite{SAM2-2025-ICLR} has provided a strong backbone for general feature representations, which was trained on 256 A100 GPUs. To reuse the pre-trained model weights and avoid prohibitive computing costs, MedSAM2 adopted the SAM2 network architecture, including an image encoder, a memory attention module, a prompt encoder, and a mask decoder (Fig~\ref{fig:1}b). The image encoder extracts multi-scale features from each 2D slice or video frame using the hierarchical vision transformer~\cite{ryali2023hiera} (Hiera), which achieves faster and more accurate performance than the na\"I've vision transformer~\cite{ViT2020ICLR} in SAM.    
The memory attention module employs transformer blocks with self-attention and cross-attention mechanisms to condition current frame features on previous frames' predictions through a streaming memory bank. 
The prompt encoders convert various user interactions (\textit{i.e.}, points, bounding boxes, and masks) to embedding. 
We used bounding boxes as the main prompt because they are less ambiguous in specifying the segmentation target, making them suitable for most organs and lesions. Specifically, for 3D images, we applied
the bounding box prompt on the middle slice and propagated the segmentation mask bidirectionally toward both ends of the volume data. Finally, the mask decoder incorporates memory-conditioned features and prompt embeddings to produce accurate segmentation masks.

Existing studies have demonstrated that fine-tuning all parts of the model yields better performance than only fine-tuning parts of the model, such as the image encoder, the mask decoder, and the prompt encoder~\cite{Microscopy-SAM,HQ-SAM}. For MedSAM2, we employ a comprehensive full-model fine-tuning approach using the lightweight SAM2.1-Tiny variant, which achieved competitive performance with fewer parameters compared to larger variants. During fine-tuning, we applied lower learning rates for the image encoder to preserve pre-trained feature extraction capabilities and higher learning rates for other model parts. We carefully balanced our training data with different sampling rates across 3D images and videos to ensure optimal performance across diverse modalities (Methods).

\begin{figure}[htbp]
\centering
\includegraphics[scale=0.43]{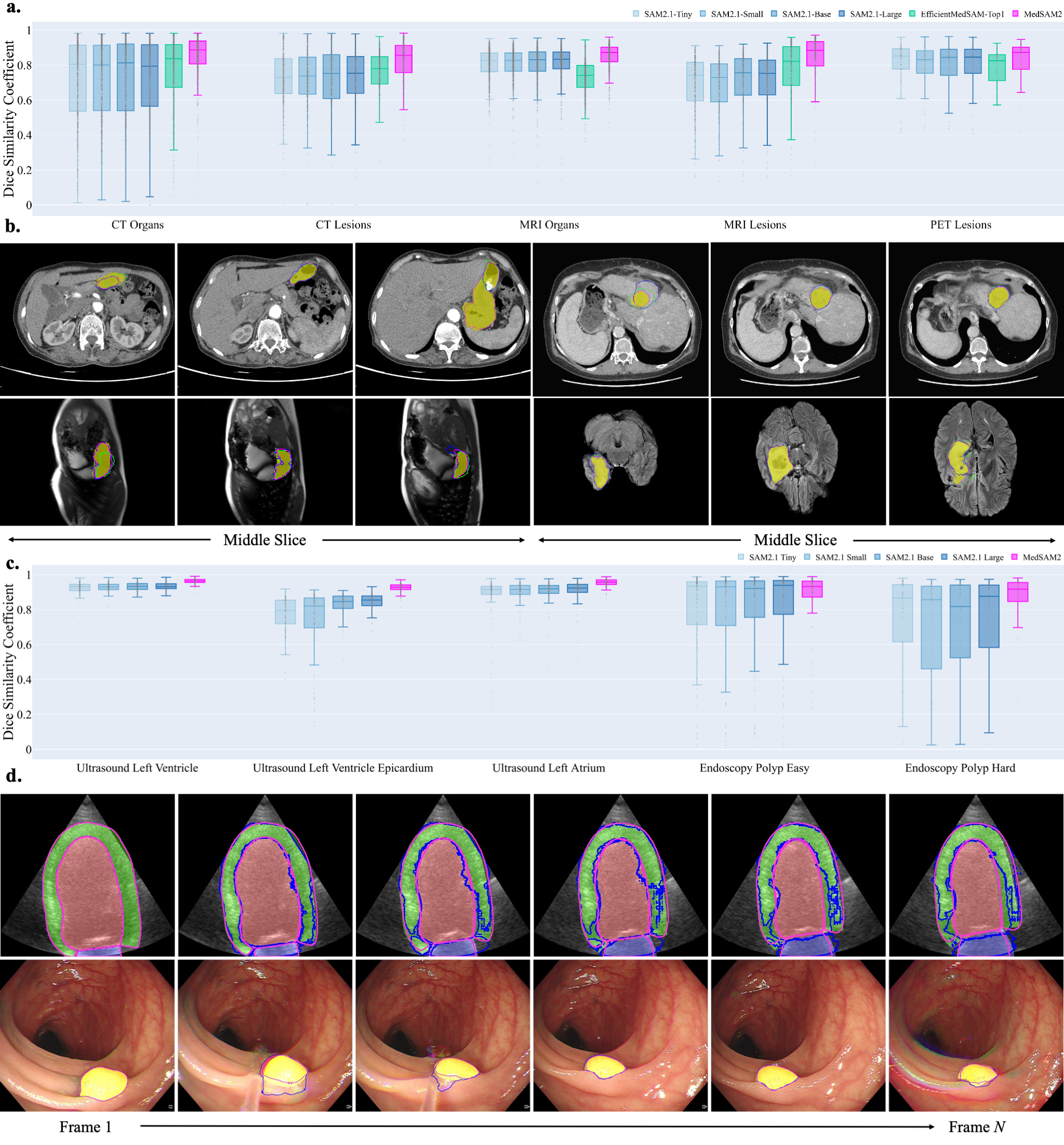}
\caption{\textbf{Segmentation performance on hold-out 3D image and video datasets.} \textbf{a,} Performance distribution of six models across five typical 3D segmentation tasks in terms of Dice similarity coefficient (DSC) scores: CT organs ($N=783$), CT Lesions ($N=409$), MRI organs ($N=734$), MRI lesions ($N=318$), and PET lesions ($N=65$). 
The center line within the box represents the median value, with the bottom and top bounds of the box delineating the 25th and 75th percentiles, respectively. Whiskers are chosen to show the 1.5 of the interquartile range. Up-triangles denote the minima and down-triangles denote the maxima. 
\textbf{b,} Visualized segmentation examples for stomach and liver cancer in computed tomography (CT), and spleen and brain cancer in Magnetic Resonance Imaging (MRI). Blue: initial bounding box prompts; Yellow: reference standards; Blue: best SAM2.1 segmentation results; Green: EfficientMedSAM-Top1 segmentation results; Magenta: MedSAM2 segmentation results. 
\textbf{c,} Performance distribution of SAM2.1 and MedSAM2 for left ventricle ($N=100$), left ventricle epicardium ($N=100$), and left atrium ($N=100$) segmentation in ultrasound videos and easy  ($N=119$) and hard  ($N=54$) polyp segmentation in endoscopy videos.  
\textbf{d,} Visualized segmentation examples for heart chambers and polyps in ultrasound and endoscopy videos, respectively. 
}\label{fig:2}
\end{figure}

\subsection*{Performance on various 3D medical image and video segmentation tasks} 
We first evaluated the trained model on the holdout 3D test set, which contains 40 segmentation tasks from different cohorts across a wide range of organs and lesions in CT, MRI, and PET scans. 
We also compared the latest SAM2.1 models with different sizes (tiny, small, base, and large)~\cite{SAM2-2025-ICLR} and the current state-of-the-art (SOTA) bounding box-based segmentation foundation model (EfficientMedSAM-Top1)~\cite{CVPR24-EffMedSAMs-2nd}, which is the winning solution in the CVPR 2024 Efficient MedSAMs competition~\cite{CVPR24-EffMedSAMs-summary}. All models were initialized with a bounding box prompt on the middle slice of the segmentation target. Each model first generated a 2D mask at the middle slice and then propagated it bidirectionally to create the full 3D segmentation.

Fig.~\ref{fig:2}a shows the quantitative results on the 3D testing set (Supplementary Table 2-3 and Fig. 1). The SAM2.1 models exhibit similar performance across all categories, with no significant differences in median DSC scores ($p-$value $>0.05$). This suggests that increasing model size within the SAM2.1 family does not necessarily translate to substantial improvements in segmentation accuracy for 3D medical images. The EfficientMedSAM-Top1 outperforms all SAM2.1 variants in CT Organs, CT Lesions, and MR Lesions, achieving median DSC scores of 83.55\% (interquartile range (IQR): 67.20-91.78\%), 77.95\% (69.15-84.81\%), and 82.25\% (68.30-90.53\%), respectively. However, its performance is not consistently superior to SAM2.1 models in MRI organ and PET lesion tasks, which is 9.22\% and 2.74\% lower than the best SAM2.1 model, respectively. One possible reason could be that the MRI Organs dataset includes images from unseen MRI sequences that introduce variations in image characteristics. 

The comparable performance of different SAM2.1 model sizes motivated us to build MedSAM2 by fine-tuning the lightweight SAM2.1-Tiny model, aiming to improve segmentation performance for medical image datasets without relying on immense computational resources.
MedSAM2 consistently achieves the highest DSC scores across all targets (CT organs: 88.84\% (80.03-94.03\%), CT lesions: 86.68\% (74.32-91.14\%), MRI organs: 87.06\% (82.96-90.04\%), MRI lesions: 88.37\% (79.91-93.26\%), PET lesions 87.22\% (79.07-90.45\%)), indicating that transfer learning is an effective way to adapt general domain foundation models to the medical image domain.
First, for simple and well-defined anatomical structures, such as the kidneys and Lungs, all methods, including SAM2.1 variants, achieve high DSC scores (often above 95\%), indicating that even general-purpose models can segment these targets accurately due to their clear boundaries and consistent appearances. However, for more challenging targets with heterogeneous appearances and complex shapes, such as the kidney lesions and pancreas, MedSAM2 shows substantial performance improvements, highlighting MedSAM2’s enhanced ability to handle greater anatomical variability.
Qualitative results (Fig.~\ref{fig:2}b) show that MedSAM2 produces more accurate and robust boundaries than other methods during propagation, owing to its memory design that effectively models temporal information across slices.

Next, we evaluated video segmentation performance for heart chambers and polyp segmentation in cardiac ultrasound (Echocardiography) and endoscopy videos on the widely used CAMUS~\cite{CAMUS} and SUN~\cite{Video-SUN-Data,Video-SUN-Seg-Gepeng} datasets, respectively (Fig.~\ref{fig:2}c, Supplementary Table 3). 
The heart chamber dataset focuses on delineating three structures: left ventricle, left ventricle epicardium, and left atrium. All SAM2.1 models perform similarly for left ventricle and atrium segmentation with high DSC scores, but have greater variance for the left ventricle epicardium in DSC score because of the heterogeneous appearances and diverse shape changes. MedSAM2 achieves better performance across the three tasks, with the highest DSC scores of 96.13\% (95.09-97.15\%), 93.10\% (91.07-94.11\%), 95.79\% (94.38-96.96\%) and less spread DSC distributions for the left ventricle, left ventricle epicardium, and left atrium, respectively, indicating better robustness in segmenting dynamic structures.

The polyp test set contains an easy and a hard subset. 
On the easy polyp subset, SAM2.1 models achieve comparable results, with similar median DSC scores ranging from 92.11\% (75.74-96.47\%) to 93.87\% (77.48-96.64\%) across different model sizes.  MedSAM2 obtains a similar median DSC score of 92.24\% (85.15-96.11\%), but exhibits a much more compact distribution with a smaller interquartile range and fewer outliers.
On the hard polyp subset, SAM2.1 models show a clear DSC score drop of 6.29\% to 10.33\% with wider variability and some outliers with low DSC scores. In contrast, MedSAM2 outperforms SAM2.1 with a noticeable gap and more consistent DSC scores of 92.22\% (83.37-95.88\%).

Qualitative segmentation results (Fig.~\ref{fig:2}d) show that SAM2.1 models struggle to capture fine structural boundaries, especially in regions with diverse contrast or complex tissue transitions. For example, the contours of SAM2.1 align with the anatomical boundaries of the left ventricle and atrium for most frames, but the segmentation quality deteriorates remarkably for the left ventricle epicardium, where the contours exhibit irregular boundaries, fragmented edges, and deviations from the true anatomical shape. MedSAM2 appears to produce smoother and more accurate segmentation results with fewer misaligned contours.
For the polyp segmentation, while all models successfully track the polyp, SAM2.1 exhibits over-segmentation by including surrounding tissues in some frames. This suggests that SAM2.1 models have difficulty maintaining spatial coherence for medical video segmentation. MedSAM2 provides a more refined and closely fitting contour, indicating its superior capability in distinguishing polyps from the background, particularly in challenging lighting and texture variations.

Altogether, SAM2.1 models perform well in simpler cases but exhibit higher variability and lower accuracy in difficult segmentation tasks. MedSAM2 consistently outperforms SAM2.1 across all tasks and produces more consistent and reliable segmentation results with reduced variability across different tasks, particularly in challenging cases, highlighting the importance of domain-specific fine-tuning for foundation models in medical image and video segmentation.

\begin{figure}[htbp]
\centering
\includegraphics[scale=0.25]{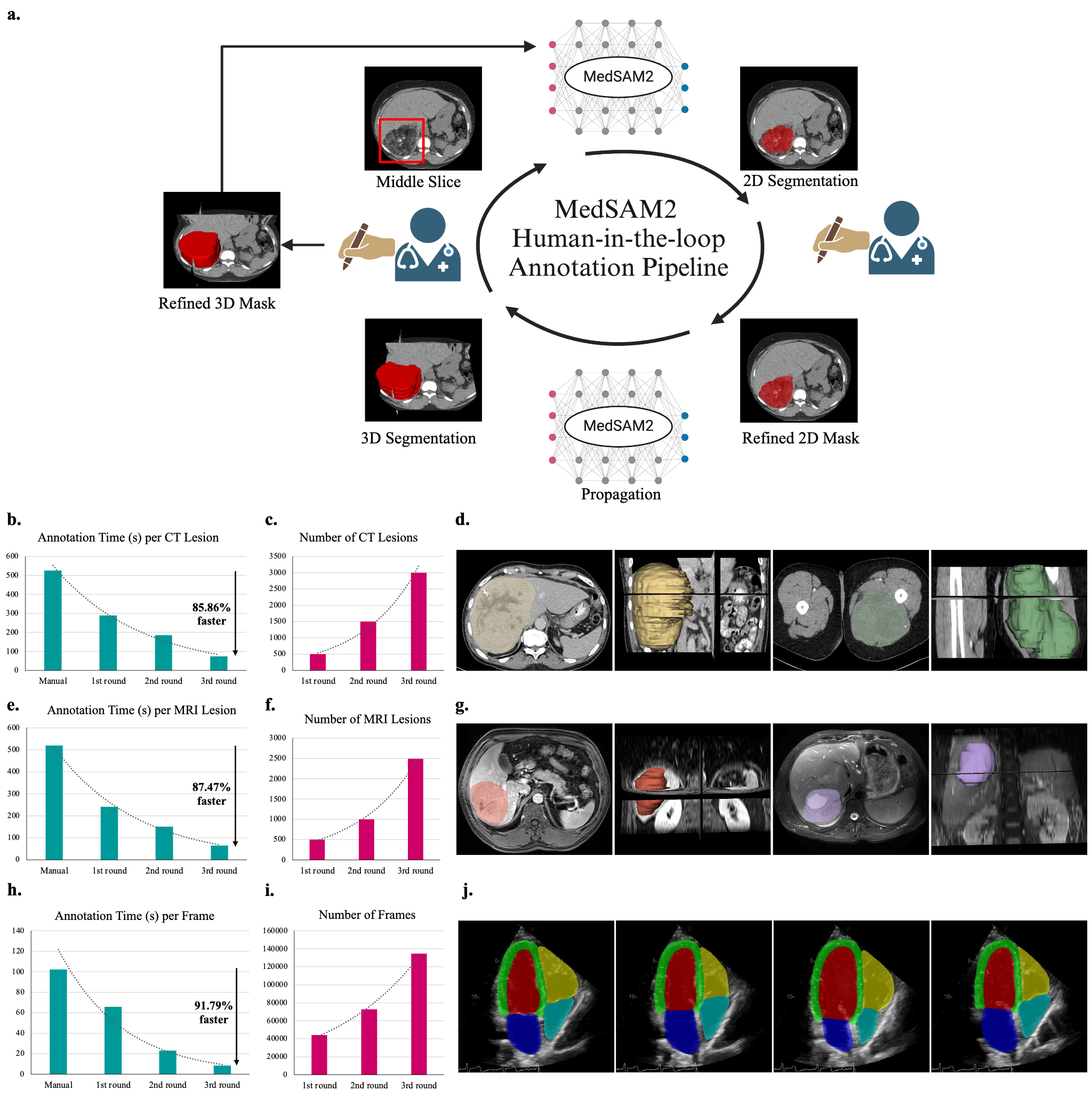}
\caption{\textbf{MedSAM2 for efficient lesion annotation in 3D CT and MRI scans.} 
\textbf{a,} A human-in-the-loop pipeline for 3D lesion segmentation. 
\textbf{b,} Annotation time per CT lesion and \textbf{c,} the number of generated CT lesions during the iterative annotation process. \textbf{d,} Visualized segmentation examples of the liver lesion and femoral osteosarcoma in CT scans. 
\textbf{e,} Annotation time per liver MRI lesion and \textbf{f,} the number of generated MRI lesions during the iterative annotation process. \textbf{g,} Visualized segmentation examples of hepatocellular carcinoma and hepatic abscess in venous contrast-enhanced phase and T2-weighted MRI scans, respectively. 
\textbf{f,} Average annotation time (seconds) per frame and \textbf{g,} the number of annotated frames during the iterative annotation process. \textbf{h,} Visualized segmentation examples of the left ventricle (red), myocardium (green), left atrium (blue), right ventricle (yellow), and right atrium (cyan).
}\label{fig:3-3DSeg}
\end{figure}

\subsection*{MedSAM2 enables efficient 3D lesion annotation for large 3D CT and MRI datasets} 
Beyond evaluating the segmentation accuracy of MedSAM2, we assess its practical value in assisting annotations of large-scale 3D lesion datasets. Accurate and efficient lesion segmentation in 3D medical images represents one of the most critical tasks for quantitative assessment of disease progression, treatment planning, and response evaluation. However, the heterogeneity of lesions (such as size, shape, texture, and contrast) and the noise and artifacts inherent in medical imags make manual segmentation a time-consuming and labor-intensive task. 

To address this limitation, we developed a human-in-the-loop pipeline with MedSAM2 to assist in 3D lesion annotation (Fig.~\ref{fig:3-3DSeg}a). Human annotators first draw a 2D bounding box, specifying the lesion at the middle slice, where the lesion usually has the longest diameter. Lesion diameter is commonly used in RECIST (Response Evaluation Criteria in Solid Tumors)~\cite{RECIST09} to measure the lesion burden in cancer therapeutics. The 2D image and the lesion bounding box are fed into MedSAM2 to generate a 2D segmentation mask followed by human revision to get the refined 2D mask. At this step, the human annotator also specifies the top and bottom slices of the lesion. Then, MedSAM2 is executed again to generate a complete 3D lesion segmentation mask by forward and backward propagating the refined mask to the top slice and bottom slice, respectively. Finally, the human annotator manually refines the 3D segmentation to obtain the accurate 3D lesion mask. When dozens of new annotations are completed, we fine-tune MedSAM2 six to fifteen epochs to get a new model with improved performance. This pipeline is iterated multiple times to generate large-scale annotations gradually. 

We first applied the annotation pipeline to lesion segmentation in CT scans. DeepLesion~\cite{DeepLesion}, the largest lesion CT dataset, was used in this study, containing a wide range of lesion types (Methods). This dataset provided 2D bounding box annotations on the key slice where the lesion reaches its maximum 2D diameter. These bounding boxes followed the RECIST guideline, which defined the lesion size with long-axis and short-axis diameter markers on the key slice. 
Our annotation pipeline runs for three iterative rounds to refine segmentation accuracy and efficiency. Fig.~\ref{fig:3-3DSeg}b-c present the average annotation time per lesion and the increasing number of annotated lesions across these rounds. 
In the first round, we selected 500 lesions of various sizes and used the trained MedSAM2 model in the annotation pipeline. Compared to manual annotation, requiring an average of 525.9 seconds per lesion, the first round reduced the annotation time by 45\%, bringing it down to 289.2 seconds per lesion.
Then we fine-tuned MedSAM2 by combining the annotated dataset and existing CT lesion cases to derive a CT lesion-specific segmentation model, which was used in the second-round annotation. Using this improved model, we annotated 1,500 additional cases, further reducing the average annotation time to 185.3 seconds per lesion.
For the third round, we updated the model again and annotated 3,000 unlabeled cases, achieving a remarkable reduction in annotation time to 74.3 seconds per lesion. 
Fig.~\ref{fig:3-3DSeg}d shows the segmentation results of two large lesions on the liver and femur. Notably, the femoral osteosarcoma was not presented in the training set, but the model was still able to generate good results, highlighting the model’s capacity to generalize to unseen lesion types.

In addition, we used the pipeline to annotate the largest multi-phase MRI liver lesion LLD-MMRI2023 dataset~\cite{LLD-MMRI}. This dataset consists of seven liver lesion types across eight MRI phases and each lesion has pre-defined bounding box prompts (Methods). Manual annotation required an average of 520.3 seconds per lesion, making it a time-intensive process. We conducted a three-round iterative annotation process similar to the CT experiments, progressively refining the segmentation model with annotated data. As shown in Fig.~\ref{fig:3-3DSeg}e-f, 
in the first round, MedSAM2 substantially reduced the annotation time by 54\% to 240.5 seconds per lesion while successfully segmenting 498 lesions. To further enhance segmentation performance, we incorporated first-round annotations into the training set and fine-tuned MedSAM2, leading to a more efficient second-round annotation process that reduced the time to 150.7 seconds per lesion and expanded the dataset to additional 996 lesions. Building on this iterative improvement, we fine-tuned MedSAM2 once more for the third round to annotate the remaining 2,490 lesions, achieving an average annotation time of 65.2 seconds per lesion. 
Fig.~\ref{fig:3-3DSeg}g visualizes two segmentation examples of different lesion types: hepatocellular carcinoma in venous contrast-enhanced MRI and hepatic abscess in T2-weighted MRI, demonstrating that the annotation pipeline effectively handles diverse lesion appearances and generalizes to multi-phase MRI images with different characteristics. Across all rounds, this iterative process enabled the annotation of 3,984 liver lesions in approximately the time it would have taken to manually annotate only 500 cases.

\subsection*{MedSAM2 enables high-throughput video annotation} 
Medical video annotation is a particularly resource-intensive and demanding task~\cite{SurgeryAI-Review} because it requires frame-by-frame labeling of anatomical structures and pathological regions, making it much more complex than static 2D image segmentation. The dynamic nature of medical videos introduces additional challenges such as motion artifacts, varying illumination, and temporal consistency. Manual annotation in such cases is tedious and expensive, making it difficult to generate sufficient labeled data for deep learning model training or large-scale studies.

We adapted our annotation strategy for video data by leveraging MedSAM2's ability to process sequential frames with spatial and temporal coherence (Supplementary Fig.2). Unlike the 3D pipeline which uses mid-slice prompting, the video pipeline begins with users adding prompts to the segmentation targets on the first frame of the video. These prompts are then passed to the pre-trained MedSAM2 model to generate initial 2D segmentation masks for each target. The human annotators then review and refine these masks to ensure high quality, followed by feeding them back into MedSAM2 for propagation, where the model extends the refined segmentation across the remaining frames. After that, users further refine the video masks as needed, ensuring an accurate delineation of the anatomical structures throughout the sequence. 
The annotated dataset is then added to the training set, allowing further fine-tuning of MedSAM2 to improve its performance on future video annotation. 

We studied the annotation pipeline for heart chamber annotation based on the right ventricular ejection (RVENet) dataset~\cite{RVENet-data,RVENet-method}, which contains apical four-chamber view cardiac ultrasound (Echocardiography) videos of 831 patients with varying image quality and heart conditions. Echocardiography is a widely used, non-invasive imaging modality for assessing cardiac function~\cite{HeartNature20,heartNature23}, offering real-time visualization of heart chambers, valve motion, and blood flow.
We applied a three-round annotation pipeline. Fig.~\ref{fig:3-3DSeg}h shows the annotation time per ultrasound (US) frame, demonstrating a substantial reduction across iterations. Manual annotation initially required 102.3 seconds per frame, whereas the first round of the pipeline reduced this time to 65.7 seconds, marking a 46\% decrease. With further refinements in the second round, annotation time dropped to 23.1 seconds, and by the third round, it reached 8.4 seconds per frame, achieving a 92\% reduction compared to manual annotation.
 
Fig.~\ref{fig:3-3DSeg}i highlights the expanding dataset size as the annotation process scales up. The first round processed 44,165 frames across 300 videos. In the second round, with the improved model, the dataset increased to 72,794 frames from 500 videos. Finally, in the third round, the pipeline annotated 134,591 frames from 1,000 videos, demonstrating its scalability and robustness. This represents a throughput increase of over 12x compared to manual annotation methods. Visualized segmentation examples are presented in Fig.~\ref{fig:3-3DSeg}j, showing that MedSAM2 accurately delineates both ventricles and atrium with consistent boundary tracking even during cardiac contraction phases.

\begin{figure}[htbp]
\centering
\includegraphics[scale=0.68]{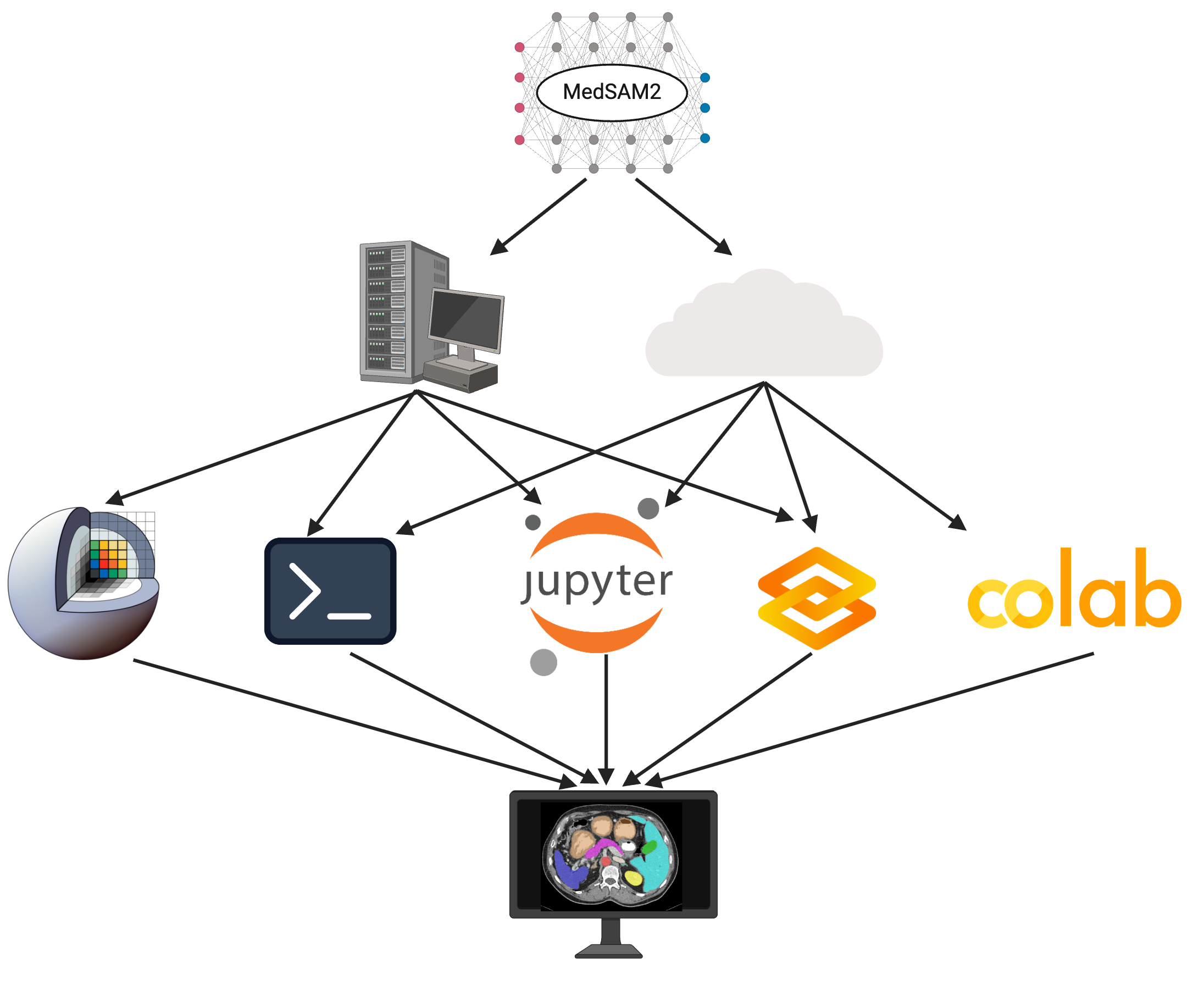}
\caption{\textbf{MedSAM2 can be deployed on local desktops and remote clusters with commonly used platforms: 3D Slicer, terminal, JupyterLab, Gradio, and Google Colab.} 
}\label{fig:deploy}
\end{figure}

\subsection*{MedSAM2 supports community-wide deployment} 
To bridge the gap between advanced segmentation models and real-world applications, we have integrated MedSAM2 into several commonly used platforms across the medical imaging and data science communities, such as 3D Slicer~\cite{Slicer}, terminal, JupyterLab, Colab, and Gradio~\cite{gradio} (Fig.~\ref{fig:deploy}). 
This multi-platform integration enables users to flexibly deploy and interact with MedSAM2 on both local desktops and remote computing environments, adapting to diverse workflows and computational resources.

3D Slicer is one of the most widely used open-source medical image analysis platforms. We implemented MedSAM2 as a plug-and-play plugin (Methods), enabling users to seamlessly apply MedSAM2 for interactive lesion and organ segmentation, visualization, and analysis in a familiar environment (Supplementary Fig. 3). This integration facilitates fast annotation and refinement of segmentation results, making it a practical tool for clinicians and biomedical researchers working with diverse 3D medical imaging modalities.

For high-throughput processing, the command-line terminal interface provides an efficient and scriptable way to process large datasets in batch mode. JupyterLab and Colab cater to researchers and developers who prefer an interactive, code-centric environment for experimentation. These platforms support notebook-based workflows, making it easy to visualize intermediate outputs, adjust model parameters, and document the segmentation process. In particular, Colab enables cloud-based access to free GPUs, allowing users without local hardware to test and deploy MedSAM2 with minimal setup.

Additionally, we incorporated MedSAM2 into Gradio, a lightweight and web-based interface that allows users to interact with the model without requiring extensive technical expertise or complex installations. This web-based deployment is particularly beneficial for video segmentation, allowing users to upload and process video frames without requiring extensive computational resources. The user-friendly design enables quick previews and adjustments of segmentations, allowing human annotators to refine results as needed. Moreover, Gradio supports seamless deployment in both local and cloud-based environments, which is essential for multi-institutional collaborations and remote research settings.

\section*{Discussion}
General segmentation foundation models, such as SAM2.1, are pre-trained on large-scale natural image and video datasets, providing strong general segmentation capabilities but typically lack the fine-grained domain knowledge required for precise medical image segmentation. 
Our results demonstrate that transfer learning is an effective strategy for adapting general-domain segmentation foundation models to medical imaging applications, enabling substantial improvements in segmentation accuracy and robustness across diverse medical imaging modalities.

Medical imaging datasets often suffer from limited annotated samples due to the high cost, time demand, and expertise required for manual annotation. Lesion segmentation is one of the most challenging tasks as they vary in size, shape, location, and contrast across different imaging modalities and patients~\cite{DeepLesion}. The scarcity of labeled data can hinder the development and generalization of general lesion detection and quantification models, limiting their clinical applicability.
Our iterative annotation pipeline with transfer learning reduced annotation time by up to 92\% while enabling dataset expansion by more than four times.

Our first user study demonstrates that fine-tuning MedSAM2 on domain-specific CT and MRI lesion datasets leads to progressive improvements in annotation efficiency and segmentation quality. 
The iterative annotation pipeline enhances the model accuracy by continuously learning from newly annotated data, reducing manual correction efforts and overall annotation time. This progressive adaptation is particularly valuable for heterogeneous datasets, such as those containing a mix of common and rare lesion types.

Video modalities, such as Echocardiography, present unique challenges compared to CT and MRI due to the dynamic nature of the heart. Unlike static medical images, ultrasound videos capture continuous motion with typical acquisition rates of 30-60 frames per second, making frame-by-frame manual annotation by experts highly impractical. This inherent complexity limits the availability of large, well-annotated segmentation datasets.
Our video annotation study demonstrates that these challenges can be effectively mitigated using an iterative annotation pipeline combined with transfer learning, achieving substantial reductions in annotation time while progressively improving segmentation quality. The model’s ability to generalize across different patient demographics and ultrasound systems further highlights its scalability.
This could further facilitate the development of cardiac assessment tools that support early disease detection and quantitative cardiac research.

Our implementation of MedSAM2 as plug-ins and packages for standard medical imaging platforms reduces adoption barriers toward translating deep learning-based segmentation models into practical tools.
By supporting deployment in 3D Slicer, terminal, JupyterLab, Colab, and Gradio, we provide both graphical interfaces and programmatic APIs for flexible access tailored to a wide range of users, from clinicians and radiologists to data scientists and algorithm developers, in both clinical research and translational settings.

This work also has several limitations. One key limitation of MedSAM2 is its reliance on bounding boxes as the main prompts. This design choice reduces object selection ambiguity and enables efficient mask propagation, allowing the model to process and track multiple masks simultaneously. However, this approach inherently limits its ability to segment highly complex anatomical structures, such as vessels with thin and branching structures. 
Since the model does not explicitly consider 3D spatial continuity, it may struggle to accurately capture highly elongated and curved 3D objects. One promising direction is the incorporation of a 4D image encoder (3D + time), which would allow the model to jointly process spatial and temporal information. Moreover, supporting other prompts, such as point~\cite{segvol,VISTA3D}, text~\cite{SAT-Yao,BiomedParse}, scribble and lasso~\cite{nnInteractive} would enable more flexible corrections.

Another limitation stems from the fixed memory design, where the model maintains an eight-frame memory bank for all segmentation tasks. While this memory size is sufficient for the majority of cases with moderate object motion, it may lead to inferior tracking performance when dealing with rapid or large target movements. For example, in colonoscopy videos, the camera continuously moves through the gastrointestinal tract, and polyps may appear, disappear, or change shape as the viewpoint shifts. Tracking failures may occur when the polyp moves out of the current memory range and then reappears in later frames. Future work will focus on implementing an adaptive memory system to replace the fixed memory bank to allocate longer memory retention for rapidly moving or intermittently visible targets.

In addition, MedSAM2 is built on the SAM2.1 tiny model with reduced input image size to optimize efficiency, but the inference process still requires GPU computation, limiting its applicability in resource-constrained environments, such as edge devices, point-of-care ultrasound machines, or low-power medical imaging workstations. Future optimizations, such as lightweight image encoder, model compression, quantization, or distillation techniques, will be necessary to enable efficient CPU-based inference, making MedSAM2 more practical in real-time and low-resource medical settings.

In conclusion, this work presents a foundation model for 3D medical image and video segmentation. We also provide, to the best of our knowledge, the most extensive user study to annotate large-scale medical datasets. 
MedSAM2 not only achieves better performance across various organs and lesions compared to existing SAM variants, but also substantially reduces annotation costs for creating large-scale segmentation datasets. 
As annotation processes become more efficient, the potential for scaling up large, high-quality labeled datasets increases, which in turn benefits future diagnostic model development and clinical deployment. Our open-source implementations across multiple platforms will facilitate adoption and further community-driven improvements to medical image and video segmentation tools.

\section*{Methods}

\subsection*{Dataset curation and pre-processing}
All training images and videos were curated from publicly available datasets with license permission for research purposes (Supplementary Table 1). The 3D test images were based on the recent 3D multi-phase liver tumor CT dataset~\cite{WAW-TACE} and the CVPR 2024 MedSAM on Laptop testing set~\cite{CVPR24-EffMedSAMs-summary}, including 20, 7, 7, 5, and 1 tasks for CT organs, CT lesions, MRI organs, MRI lesions, and PET lesions, respectively. 
The pre-processing followed common practice~\cite{nnunet21,MedSAM,BiomedParse}. 
Specifically, CT image intensities were adjusted to the proper window width and level (brain: 80/40, abdomen: 400/40, bone: 1800/400, lung: 1500/-600, mediastinum: 400/40) followed by rescaling to $[0, 255]$. For the remaining 3D images (MRI and PET), we applied an intensity cut-off with a lower-bound and upper-bound of 0.5\% and 99.5\% percentile of foreground intensity and then rescaled the intensity to $[0, 255]$. No intensity normalization was applied for videos.

\subsection*{Lesion CT dataset and annotation pipeline}
DeepLesion dataset~\cite{DeepLesion} contains 32,735 diverse lesions in 32,120 CT slices from 10,594 studies of 4,427 unique patients. Each lesion has a bounding box annotation on the key slice, which is derived from the longest diameter and longest perpendicular diameter. 
We prioritized lesions with a minimal diameter of $25 mm$ because larger lesions are more time-consuming during manual annotation. A senior radiologist with more than 10 years of experience manually annotated five cases to get the manual annotation time. In the human-in-the-loop experiment, we first generated the 2D segmentation mask on the key slice by MedSAM2 and then two radiology students manually revised the mask and specified the top slice and bottom slice of the lesion. To improve the efficiency, we concatenated eight preprocessed lesion images along the axial plane as one 3D volume. In this way, human annotators can open eight lesion images at once for manual revision and reduce the time costs to adjust the window level and width for each lesion. 
All lesion images and masks were resampled to $512\times512$ on the axial plane in the concatenation with third-order spline interpolation and nearest-neighbor interpolation, respectively. 
Images with out-of-the-plane spacing less than $3mm$ were resampled to $3mm$. After manual revisions, we separated the merged eight-lesion scan into single images and resampled them to the original shape. 
We excluded images without measurable lesions or an out-of-plane spacing of more than $5mm$. 
Finally, all annotations were checked and revised by the senior radiologist.

\subsection*{Liver lesion MRI dataset and annotation pipeline}
LLD-MMRI dataset~\cite{LLD-MMRI} contains diverse liver lesions from 498 unique patients, including hepatocellular carcinoma, intrahepatic cholangiocarcinoma, liver metastases (HM), hepatic cysts (HC), hepatic hemangioma, focal nodular hyperplasia, and hepatic abscess. Each lesion has eight MRI scans: non-contrast, arterial, venous, delay, T2-weighted imaging, diffusion-weighted imaging, T1 in-phase, and T1 out-of-phase, resulting in 3984 cases in total. Each liver lesion has both 3D and slice-wise 2D bounding boxes. 
We ran MedSAM2 with the two types of bounding boxes separately and got two groups of segmentation results. For the 3D bounding box prompt, we first generated a 2D segmentation mask on the median slice followed by propagating the mask to the remaining slices until it reached the top and bottom slices. 
For the 2D bounding box prompt, we ran MedSAM2 for each slice with the corresponding box prompts. 
After that, we computed the DSC score between the two groups of segmentation results. We hypothesized that hard cases have larger disagreements between the two segmentation masks. For each patient, we selected the case with the lowest DSC score among the eight MRI scans as the first-round revision candidates, aiming to achieve a trade-off between data diversity and difficulty. The same selection criteria were also used in the second-round iteration. 
A senior radiologist manually annotated five cases to get the manual annotation time. 
Two radiology students participated in the manual revision process. Different from the CT lesion annotation, slice-wise 2D bounding box-based segmentation results were used in the revision because we found the segmentation accuracy was better than the 3D bounding box-based results. 
During the revision process, we resampled the images to $352\times352$ and merged five preprocessed lesion images as one volume for better efficiency. Finally, all annotations were checked and revised by the senior radiologist.

\subsection*{Cardiac ultrasound (echocardiography) video dataset and annotation pipeline}
RVENet dataset~\cite{RVENet-data,RVENet-method} consists of 3583 echocardiography videos from 831 unique patients. The same annotation protocol in CAMUS dataset~\cite{CAMUS} was followed to delineate the left ventricle, myocardium, and left atrium. Since the videos were acquired in the apical four-chamber view, the right ventricle and atrium were also annotated to provide a more comprehensive cardiac analysis. 
Videos with low image quality or incomplete ventricles and atrium were excluded. The raw videos have a high resolution of $1016\times708$ and $800\times600$. We downsampled the videos by a factor of two to reduce the annotation workload while essential structure details were preserved to differentiate the heart chambers. 
We first annotated the first frame of 200 videos from different patients with the bounding box or point prompts followed by manual refinement by three radiology students. 
The corrected first-frame mask was then propagated across subsequent frames using MedSAM2. To enhance segmentation accuracy, human annotators manually refined three to ten frames at approximately uniform intervals before inferencing MedSAM2 to update the segmentation results. Finally, all frames underwent manual adjustments where necessary, and the annotations were rigorously verified by the senior radiologist before being used to fine-tune MedSAM2 for the next iteration.
To compare with manual annotation efficiency, a senior radiologist annotated 10 frames as a reference for manual annotation time cost.

\subsection*{Network architecture}
MedSAM2 was built upon SAM2~\cite{SAM2-2025-ICLR} with four main components: an image encoder, a prompt encoder, a memory attention module, and a mask decoder.
First, we modified the image encoder by downsizing the input image size from $3\times1024\times1024$ to $3\times512\times512$, which not only fitted better for typical medical image size but also reduced computational burden. The image encoder employs a hierarchical vision transformer (Hiera)~\cite{ryali2023hiera} with a four-stage architecture (layers=\{1,2,7,2\}). We incorporated global attention blocks at the 5th, 7th and 9th layers to capture long-range dependencies critical for medical image analysis. A feature pyramid network (FPN)~\cite{lin2017feature} neck extracts multi-scale features from the backbone, enabling detailed segmentation at various resolutions. 
Second, the memory attention module contains 4 transformer layers with both self-attention and cross-attention mechanisms. Each layer employs Rotary Position Embedding (RoPE)~\cite{su2024roformer} with 2D spatial encoding (feature size 32$\times$32) to maintain spatial awareness across slices or frames. This module conditions the current frame features on a memory bank storing information from previously processed frames, effectively exploiting the spatial continuity in volumetric data and temporal coherence in videos. 
Third, the prompt encoder transforms coordinates into embeddings that guide the segmentation process, allowing clinicians to specify regions of interest efficiently. Finally, the mask decoder integrates features from multiple scales of the image encoder through skip connections and produces segmentation masks at 128$\times$128 resolution, which are then upsampled to the original 512$\times$512 input size using bilinear interpolation.

\subsection*{Training protocol}
The model was initialized from the pre-trained SAM2.1-Tiny model checkpoint. 
During training, we used a full model fine-tuning strategy with two different learning rates: a lower learning rate ($3.0 \times 10^{-5}$) for the image encoder (28M parameters) to preserve learned features, and a higher rate ($5.0 \times 10^{-5}$) for other components (10.9M parameters) to adapt to the characteristics of the medical domains. 
The training utilized a combination of 3D images and videos with a batch size of eight per GPU, where each training sample consisted of eight consecutive slices or video frames. In the human-in-the-loop annotation study, we halved the learning rate and fine-tuned the trained MedSAM2 model 6 and 15 epochs in the second and third round iterations, respectively.

The data augmentations included random horizontal flipping, affine transformations, color jittering, and random grayscale conversion. For videos, we also augmented the frame sample rate by a factor of 2 and 4.  
Since the training set was imbalanced between different modalities, we increased the sampling frequency of MRI, PET, and video data by a factor of 3, 40, and 40, respectively. 
The bounding box prompts were simulated from expert annotations with random perturbations of 0-10 pixels. The loss function combined focal loss and dice loss for mask prediction with weights of 20:1. We used the AdamW optimizer~\cite{adamW} with $\beta_1 = 0.9$, $\beta_2 = 0.999$, and weight decay of 0.01. The model was trained for 70 epochs on three compute nodes, each equipped with four H100 GPUs, with a total training time of four days. External validation was performed on held-out datasets to assess the model's generalization capability across different tasks and modalities.

\subsection*{3D Slicer Integration}
We implemented MedSAM2 as a plugin (extension) in 3D Slicer to reuse the built-in modules for essential operations, such as loading diverse medical imaging formats (\textit{e.g.}, DICOM, NIfTI), drawing prompts, refining masks, and visualizing both 2D slices and 3D segmentation results. 
The plugin is built on a client-server architecture, offering users the flexibility to perform inference either locally on personal machines or remotely on high-performance computing clusters.
The interface contains three clear sections: 
\begin{itemize}[leftmargin=*]
    \item Preprocessing panel: users can select predefined pre-processing options (\textit{e.g.}, CT, MRI) to normalize the input image intensity before segmentation. 
    \item Region-Of-Interest (ROI) selection: users can define the ROI directly by choosing start and end slices and draw bounding boxes prompts on the key slice. 
    \item Segmentation controls: users can choose the model variant and initiate segmentation for the middle slice and full volume. Moreover, users can load their own customized models for specific imaging modalities or segmentation targets. 
\end{itemize}
For the server component, we implemented a Flask API server to provide the necessary arguments and inputs to the local API offered by MedSAM2. The server also features a temporary most recently used (MRU)-style cache to facilitate refinement of the most recent segmentation.

\subsection*{Evaluation metrics and platform}
We followed the recommendations in Metrics Reloaded~\cite{metric-reload} to evaluate the segmentation accuracy. Specifically, we used Dice Similarity Coefficient (DSC) and Normalized Surface Distance (NSD) with a boundary tolerance of $2 mm$ to quantitatively evaluate the region overlap and boundary similarity, respectively. For CT, MRI, and PET images, we compute the metrics in 3D while for video datasets, we first compute the frame-wise metric scores followed by averaging them to obtain the video-level metric scores. Wilcoxon signed-rank test was used for statistical significance analysis. Results were considered statistically significant if the $p-$value was less than 0.05.

\subsection*{Data availability}
All data used in the study are from public datasets, and detailed references are provided in Supplementary Table 1. We also create a dedicated website, accessible at \url{https://medsam-datasetlist.github.io}, to provide a comprehensive and continuously updated repository of medical image segmentation datasets. This resource is intended for long-term maintenance and accessibility to the research community.

\subsection*{Code availability}
The code, model weights, and annotated datasets are publicly available at \url{https://github.com/bowang-lab/MedSAM2}. 3D slicer plugin can be accessed at
\url{https://github.com/bowang-lab/MedSAMSlicer}.

\bibliographystyle{IEEEtran}
\bibliography{main-ref}

\end{document}